%% file: main.tex
\documentclass[conference]{IEEEtran}
\IEEEoverridecommandlockouts
\usepackage{cite}
\usepackage{amsmath,amssymb,amsfonts}
\usepackage{textcomp}
\usepackage{xcolor}
\def\BibTeX{{\rm B\kern-.05em{\sc i\kern-.025em b}\kern-.08em
    T\kern-.1667em\lower.7ex\hbox{E}\kern-.125emX}}
\usepackage{times}
\usepackage{amssymb}
\usepackage{url}
\usepackage{hyperref}

\usepackage{booktabs} % For formal tables
\usepackage{algorithmic}
\usepackage{algorithm}
\usepackage{longtable}
\include{usepackages}
\include{newcommands}

\usepackage{authblk}

\usepackage[colorinlistoftodos]{todonotes}
\usepackage{subcaption}

\usepackage{caption}
\usepackage{wrapfig}

\begin{document}

\title{Secure Integration of Electric Vehicles with the Power Grid}

\author[*]{Chaitra Niddodi}
\author[+]{Shanny Lin}
\author[*]{Sibin Mohan}
\author[+]{Hao Zhu}
\affil[*]{University of Illinois at Urbana-Champaign}
\affil[+]{University of Texas at Austin}

\maketitle
\pagestyle{plain}
\input{abs}
\input{intro}
\input{bg}
\input{system}
\input{threat}
\input{ids}
\input{impl}
\input{eval}
\input{disc}
\input{rwork}
\input{concl}
\input{ack}

%\Urlmuskip=0mu plus 1mu
\bibliographystyle{plainurl}
%\bibliography{main}
%\input{main.bbl}
\bibliography{refs}

\input{app}
\end{document}

%% file: usepackages.tex
\usepackage[inline]{enumitem}

\usepackage[utf8]{inputenc}
\usepackage{amsthm} % for proof
\usepackage{amsmath}
\usepackage{graphicx}
\usepackage{boxedminipage}
\usepackage{epsfig}
\usepackage{epstopdf}
\usepackage{psfrag}
\usepackage{comment}
\usepackage{afterpage}
\usepackage{pstricks}
\usepackage{graphics}

\usepackage{mdwtab}
\usepackage{wrapfig}
\usepackage{graphics}

\usepackage{algorithm}  
\usepackage{algorithmic} 
\usepackage{amssymb} % for \triangleq denotation in the appendix

\usepackage{makecell}

\usepackage{hhline}

\usepackage{tabularx}

\usepackage{hyperref}

\usepackage{epsfig}
\usepackage{fixfoot}
\usepackage{listings}
\usepackage{psfrag,wrapfig}
\usepackage{xspace,soul}
\usepackage[colorinlistoftodos]{todonotes}
\usepackage{subcaption}

%% file: newcommands.tex
\newcommand{\eg}{{\it e.g.,}\xspace}
\newcommand{\viz}{{\it viz.,}}

\newcommand{\ie}{{\it i.e.,}\xspace}

\newcommand{\etc}{{\it etc.~}}
\newcommand{\ci}{{\it (i) }}
\newcommand{\cii}{{\it (ii) }}

\newcommand{\ca}{{\it (a) }}
\newcommand{\cb}{{\it (b) }}
\newcommand{\cc}{{\it (c) }}
\newcommand{\cd}{{\it (d) }}
\newcommand{\ce}{{\it (e) }}

%% file: abs.tex
\begin{abstract}
    
    This paper focuses on the secure integration of distributed energy resources (DERs), especially pluggable electric vehicles (EVs), with the power grid. We consider the vehicle-to-grid (V2G) system where EVs are connected to the power grid through an 'aggregator'\footnote{Aggregator is the component in the V2G system that connects EVs to the power grid (Fig. \ref{fig::grid_overview})}. In this paper, we propose a novel \textit{Cyber-Physical Anomaly Detection Engine} that monitors system behavior and detects anomalies almost instantaneously (worst case inspection time for a packet is 0.165 seconds\footnote{Minimum latency on V2G network is 2 seconds}). This detection engine ensures that the critical power grid component (\viz aggregator) remains secure by monitoring \ca cyber messages for various state changes and \textit{data constraints} along with \cb power data on the V2G cyber network using power measurements from sensors on the physical/power distribution network. Since the V2G system is time-sensitive, the anomaly detection engine also monitors the \textit{timing requirements} of the protocol messages to enhance the safety of the aggregator. To the best of our knowledge, this is the first piece of work that combines \ca the EV charging/discharging protocols, the \cb cyber network and \cc power measurements from physical network to detect intrusions in the EV to power grid system.
    
\end{abstract}


%% file: intro.tex
\section{Introduction}
\label{sec::intro}

The power grid is one of the critical infrastructures of a nation. It
is a complex cyber-physical system (CPS)
%\footnote{Cyber-physical systems (CPS) are engineered systems that are built by integrating computation components with physical components. \cite{cps}.}, 
often with time sensitive properties. 
Any problem that affects the power grid can result in damage to life, property or the environment. 
With increased automation, technological and communication advances, multiple new components
and systems are interacting with the grid. Some notable
examples are the distributed energy resources (DERs) such as pluggable
electric vehicles (EVs), solar arrays, smart homes and industrial building
automation systems. With this increased connectivity
and plethora of applications comes more opportunities for malicious
entities to gain access to critical systems and potentially wreak havoc
with essential components of the power grid. 
%For instance, an attacker could
%enter the system through a vulnerability in the control code or 
%communication channel for a building automation system and then fluctuate
%(either by making actual changes or faking it) the power usage and
%requirements for a large industrial building. This could result in 
%variations in the local grid that could percolate to larger sections of 
%the system. Given enough time and by taking control of a sufficient number
%of such DERs, attackers could cause the grid to become unstable, even
%leading to a crash. 
Attackers could use the communication channels between the DERs and the 
power grid to potentially take control of or even shut down critical grid 
components 
% \cite{ukraine-attack}. \hl{Was this ukraine attack an example of
% attackers using comm. between DERs and grid? - yes there were DoS attacks launched}

There exist multiple challenges in securing power grid systems: 
\ca the grid has many legacy systems that may not know how to interact, 
in a secure manner, with the newer applications such as EVs;
\cb attacks on edge devices (such as DERs) are difficult to detect at
the grid level due to the lack of visibility into their operational details;
\cc new systems such as the Vehicle-to-Grid (V2G) introduce new components such as EVs,
aggregators and electric vehicle supply equipment (EVSEs) to the grid, each
of which can have additional vulnerabilities;
\cd these systems also introduce new communication paths that raise new
issues dealing with coordination among multiple stakeholders and finally
\ce new infrastructure must be set up for managing credentials for the
new components and their operators. In fact, it is well documented that
the power grid is vulnerable to a wide range of attacks \cite{nes}.

With the increasing number of EVs (\eg Tesla), the
communication complexities in the V2G system have increased.
EVs not only participate in drawing current from the grid to charge themselves
but they also act as batteries that can discharge to the grid during periods of high power demand \cite{agg2}. Hence, EVs are fast becoming an important type of DER
that closely interact with the grid. 
% If an attacker can take control of enough umber of EVs in a local area and in turn take control of the entity that controls/manages a
% large number of them (\viz aggregator), then these systems can be used to cause problems to the power grid.
% Hence, it is vital to ensure that EVs (and especially their communication
% channels with the power grid) are monitored and remain secure. In this work we
% focus on EVs and their connection to the grid via intermediaries such as 
% aggregators.

\begin{figure}
  \begin{center}
    \includegraphics[scale=0.35]{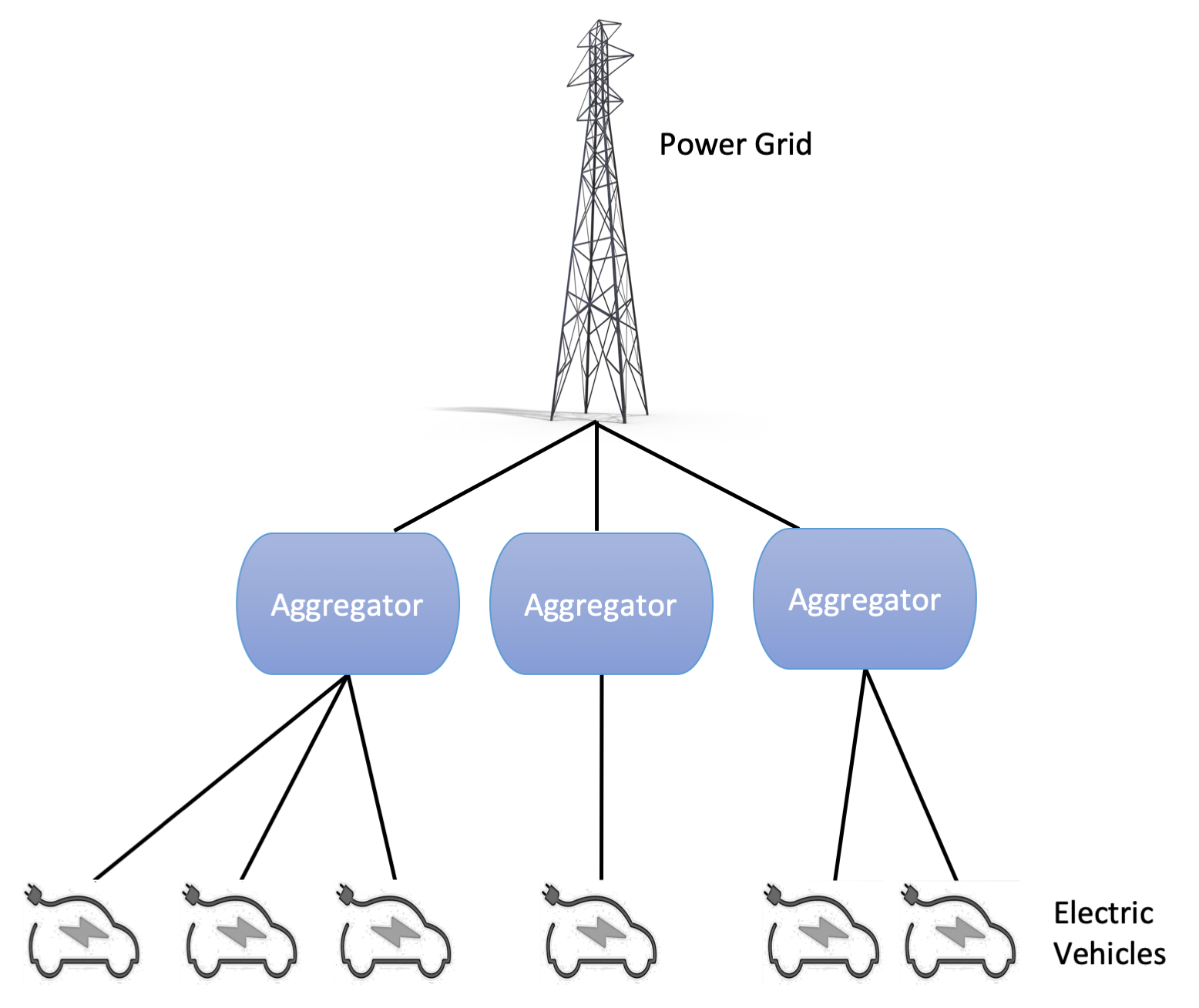}
  \end{center}
  \caption{Conceptual Architecture of a Vehicle-to-Grid (V2G)system.\cite{wip}}
  \label{fig::grid_overview}
\end{figure}
Figure \ref{fig::grid_overview} presents a high-level conceptual architecture
of the power grid with multiple EVs connected to it. The ``aggregators'' \cite{agg} in this
figure are the entities that act as mediators between the end DER systems 
(the EVs in this case) and the power grid utility system. A single aggregator
can manage multiple EVs (usually in close geographic proximity). In the model presented, the aggregator
can be a prime target for attackers since \ci it manages multiple EVs
and \cii is also directly connected to the power grid utility system. A successful
intrusion at the aggregator level can have serious consequences for the
power grid. Therefore, ensuring the security of this critical component (\viz aggregator) is essential to ensure secure integration of DERs such as EVs with the power grid.
To this end, we {\em propose a Cyber-Physical Anomaly Detection Engine} 
with mechanisms to detect anomalous behavior in aggregators of the V2G system. 
%As such, we focus on the following:
%\ca the security of the aggregators, \cb the communication between the
%aggregators and the DERs (specifically, EVs in our case) and 
%\cc the connections between the aggregators and the grid. 
For our anomaly detection engine, we rely on both the cyber and
physical properties of the system. On the cyber side, we focus on the
communication protocol in the V2G system to ensure correct operation
of the aggregator, while we validate its behavior using the physical 
side of the system in the form of power measurements.\\

\noindent The main \textbf{contributions} of this work are:
\begin{enumerate}
	\item An enumeration of the correct sequences of commands in the V2G 
	communication protocol. This is used to generate an aggregator state machine for 
	our detection engine (\S{\ref{subsec::ms}})
	\item Development of a Cyber-Physical Anomaly Detection Engine that can detect malicious activity at the aggregator level 
	as soon as they occur. The anomaly detection engine monitors communication on the V2G cyber network using power measurements from sensors on the physical/power distribution network (\S{\ref{subsec::cp}}). It also 
	makes use of \textit{timing constraints} related to frequency of periodic messages (\S{\ref{subsec::mf}}) and subscription period (\S{\ref{subsec::sp}})
	%, along with data constraints, 
	to differentiate between correct/incorrect system behavior.
	\item Implementation and evaluation of a prototype of the 
	Cyber-Physical Anomaly Detection Engine (\S{\ref{sec::impl} and \S{\ref{sec::eval}}})
\end{enumerate}
While there are some intrusion detection systems designed for components of the power grid system (as discussed in detail in \S{\ref{sec::rel_work}}), to the best of our knowledge there are none that combine cyber and physical properties of the system along with communication standards for EVs. Hence, there is no direct comparison possible while evaluating our cyber-physical anomaly detection engine. Our evaluation consists of measuring the accuracy and performance of our anomaly detection engine as described in \S{\ref{sec::eval}}. 
The simple model of our anomaly detection engine enables it to detect anomalies accurately and almost instantaneously.

% %\FloatBarrier
% \begin{figure}[h]
% 	\centerline{\includegraphics[scale=0.35]{images/v2g.png}}
% 	\caption{Conceptual Architecture of a Vehicle-to-Grid (V2G) system.\cite{wip}}
% 	\label{fig::grid_overview}
% \end{figure}
% %\FloatBarrier

%% file: bg.tex
%\newpage
%\setcounter{figure}{0}
%\setcounter{table}{0}
\section{Vehicle-To-Grid System}
\label{sec::bg}
%\subsection{Vehicle-to-Grid System Model}
%\label{bg1}

Fig.\ref{fig::grid_overview} shows the conceptual architecture of the V2G system. The main components in this system include EVs, aggregators and the power grid. 
EVs not only act as loads but can also feed power back to the grid. The EV to power grid operations considered in this paper are \ca  charging where EV draws power from the grid, \cb discharging where EV supplies power to the grid during times of peak power demand and helps to reduce the load on the grid 
% and \cc frequency regulation, an operation that is used to stabilize the grid frequency by performing frequent charging and discharging operations. 

The role of aggregators as mediators between end users (\viz EVs) and the utility operator is particularly useful in coordinating discharging 
% and frequency regulation 
operation.
This is because individual EVs have very small power capacities in comparison with the scales of power generation and distribution at the power grid. Therefore, for efficient discharging 
% and frequency regulation
operations, a large number of EVs are required. An aggregator manages multiple EVs and helps in efficiently managing these operations \cite{agg2}. With aggregators acting as intermediaries between the utility power grid operator and the EVs \cite{agg}, all communication messages between the EVs and the power grid pass through aggregators.
%This is the system model used in our work.

The requirements and specifications for communication between the
EVs and the electric power grid are established by the SAE communication standards \cite{sae}. SAE J2847/1 standard provides specifications for forward power flow \viz charging operation between the EVs and the power grid \cite{sae}. Whereas, SAE J2847/3 standard provides specifications for reverse power flow \viz discharging 
% and frequency regulation
operation between the EVs and the power grid \cite{sae}.

Integration of distributed sources of energy such as EVs
with the power grid comes with the penalty of making the grid
susceptible to a range of cyber-physical attacks. These include large
scale attacks if many of these edge devices \viz EVs, are compromised. The growing number of sophisticated attacks (\S \ref{sec::threat}), necessitates the need for the \textit{development of an advanced cyber-physical attack detection and resiliency framework}.
% The vulnerability of the V2G system to
% a wide range of attacks is mainly due to the following factors:
% \begin{itemize}
% 	\item Utilities do not have direct control over EVs and hence are
% 	unable to enforce strict policies for secure communication.
% 	\item Customer sites lack enough security. This leads
% 	to exposure of EVs to physical attacks such as attacks
% 	on locks and other anti-theft mechanisms \cite{aut}. There is also an additional risk of exposure to
% 	cyber attacks due to insufficient knowledge, for instance,
% 	in setting up passwords.
% 	\item Direct interaction of customers with EVs may be used to
% 	unethically manipulate energy consumption data \cite{ev_false}. Compromised
% 	grid edge devices (EVs) may also be used to
% 	launch attacks that cause grid instabilities and blackouts \cite{ukraine-attack}.
% \end{itemize}

%% file: system.tex
%\newpage
%\setcounter{figure}{0}
%\setcounter{table}{0}
\section{System Model}
\label{sec::system}

% We use the model from literature \cite{agg} where aggregators act as intermediaries between the utility (power grid) operator and the EVs.
% Hence all communication messages between the EVs and the power grid in the V2G system, pass through the aggregators. To ensure safe operation of the aggregator, we develop a Cyber-Physical Anomaly Detection Engine that resides at the aggregator.
%\FloatBarrier
\begin{figure}[h]
	\centerline{\includegraphics[scale=0.24]{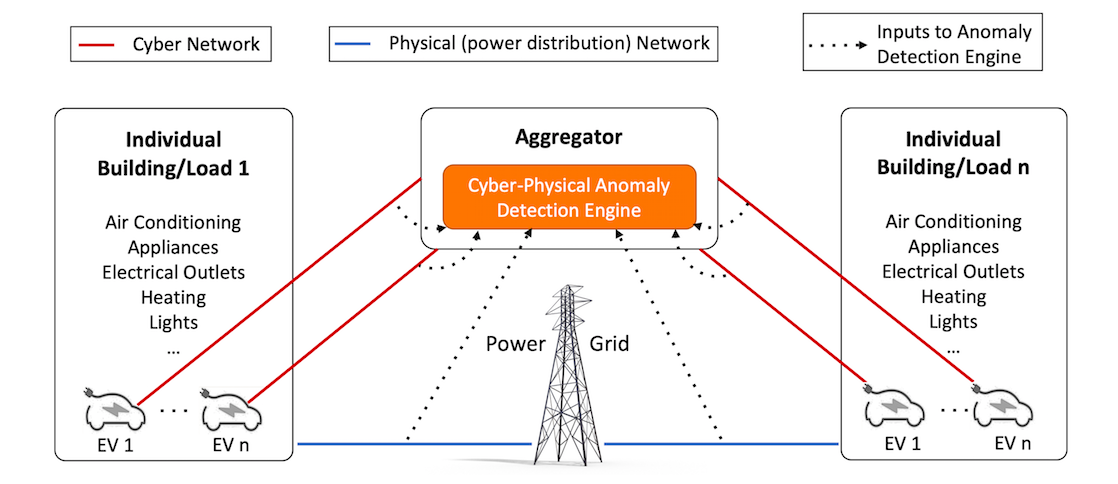}}
	\caption{System Model: EVs intermittently charging at homes}
	\label{fig::sysmodel}
\end{figure}
%\FloatBarrier

Fig.\ref{fig::sysmodel} highlights the system under consideration along with the various components and relevant connections in the cyber and physical networks. Multiple buildings (\eg households) are connected to the aggregator with multiple EVs connected to each building. The Cyber-Physical Anomaly Detection Engine residing at the aggregator receives inputs from both the cyber as well as physical networks. We assume a second cyber network connecting the sensors in the physical network / power distribution network to the Cyber-Physical Anomaly Detection Engine in order to receive power measurements. For the purpose of evaluating the performance of our system, we assume that this second cyber network has similar properties (in terms of bandwidth etc) as the cyber network in the V2G system. 

As part of the cyber network, the EVs are the only type of load being considered, i.e, only the packets exchanged between the EVs and aggregator are taken as input for anomaly detection. As part of the physical network, the aggregated profile of all household loads at the entry point
%\todo[inline]{@shanny: what types of load does it include (updated)} 
is taken as input for anomaly detection. %detecting EV charging activities. 
Individual household loads %are taken into consideration as the aggregated load. 
such as air conditioning, appliances, electrical outlets, heating and lighting loads etc. make up the majority of the non-EV loads in the aggregated profile. 
%From this aggregated load profile, the EV-only component, captured by its start/end charging patterns of fast and large power changes, can be differentiated from other non-EV household loads. This filtered out EV-only power obtained from the physical network is cross-validated against the EV power obtained from the cyber network in the V2G system.
In this model, we assume that the physical network and cyber network are not compromised simultaneously, \ie we assume that the inputs obtained from atleast one of these networks are genuine.

%% file: threat.tex
%\newpage
%\setcounter{figure}{0}
%\setcounter{table}{0}
\section{Threat Model}
\label{sec::threat}

% Integrating EVs (that act as variable loads) with the power
% grid entails many security considerations. 

%Tab. \ref{tab1} shows the possible attacks on a V2G system \cite{t1,t2,t3}. 
With reference to the power grid, security issues manifest as safety/reliability concerns where attackers try to bring down system reliability. The attacks in a V2G system can be broadly classified into Network-based attacks and Component-based attacks. 
\textit{Network-based attacks} are the ones where the V2G communication network is compromised. 
These include Man-in-the-middle attacks (MITM) that either passively intercept and observe the communication by eavesdropping or actively modify communication by injection, replay or repudiation of traffic. Denial of Service (DoS) attacks on the V2G network include jamming of signals and dropping of packets. 
For instance, when EVs initially connect to the grid, information such as customer details and location data are
exchanged. Eavesdropping on such information compromises
customer privacy. 
As another example, transmitted control commands and updating
firmware, software, drivers \etc also affect system stability,
safety and reliability. Active modification of such traffic has an impact
on the functioning of the V2G components \cite{v2g}. 
\textit{Component-based attacks} are those attacks where one or more of the components in the V2G system are compromised. 
These include violation of authentication and/or authorization at the utility system (components of the power grid including aggregators), spoofing of utility system components, compromise of end devices (Electric Vehicles) and Denial of Service (DoS) attacks on utility system components by the exhaustion of resources at aggregator and/or Power Grid. 
These threats make the V2G
sub-system of the power grid highly susceptible to attacks.

%\FloatBarrier
\begin{table}[h]
	\centering
	\caption{Threat Model}\label{tab::threat}
	\renewcommand{\arraystretch}{1}
	\begin{tabular}{|p{0.25cm}|p{4cm}|p{3cm}|}
		\hline
		\textbf{No.} & \textbf{Attack} & \textbf{Effect} \\ 
		%& \textbf{Detection Mechanism used by Anomaly Detection Engine}\\
		\hline
		1 & Compromise of EVs or the cyber network to report more power than actually consumed  & DOS attacks at the aggregator preventing more EVs from connecting to aggregator \\ 
		%& Validating consistency of cyber and physical states  \\
		\hline
		2 & Compromise of EVs or the cyber network to report less power than actually consumed  & Could cause transformer overheating due to more EVs connecting to aggregator \\ 
		%& Validating consistency of cyber and physical states  \\
		\hline
		3 & Compromise of EVs, the cyber network or the power network causing packets to be generated out of expected sequence  & Disrupt the correct functioning of aggregator to cause aggregator instabilities \\ 
		%& Message sequence validation  \\
		\hline
		4 & Compromise of EVs, the cyber network or the power network causing EVs to charge/discharge beyond their subscription periods  & Disrupt the correct functioning of aggregator to cause aggregator instabilities \\ 
		%& Subscription period validation  \\
		\hline
		5 & Compromise of EVs, the cyber network or the power network causing periodic message packets to be generated more or less frequently than expected  & Disrupt the correct functioning of aggregator to cause aggregator instabilities \\ 
		%& Frequency validation of periodic messages  \\
		\hline
	\end{tabular}
\end{table}
%\FloatBarrier

Based on the above possible attacks in a V2G system \cite{t1,t2,v2g}, 
Tab.\ref{tab::threat} shows the attacks that our Cyber-Physical Anomaly Detection engine focuses on. There have been instances of attacks involving modification of current to cause a fire \cite{evat1}.
Multiple studies \cite{w1,w2} have shown that EVs can offer significant services to improve grid stability. It therefore follows that compromising a large number of EVs and in turn a large number of aggregators using the above attacks (Tab. \ref{tab::threat}) could cause grid instabilities. Hence securing the component that connects EVs to the power grid \viz aggregator is of paramount importance for
the safe operation of the power grid.

%% file: ids.tex
\section{Cyber-Physical Anomaly Detection Engine}
\label{sec::ids}

%Power grid systems are often time-sensitive. Therefore, our goal is to not only detect anomalies accurately but to also do so at a rate that avoids introducing significant processing delay into the system.

%In case of an anomaly, we either notify the power grid or switch control to a trusted aggregator, as explained in detail below.

%\subsection{Assumptions}
%%\indent % indent 1st para also
%
%\begin{enumerate}
%	\item We assume that the V2G system is clean and not compromised when the intrusion detection system starts monitoring the aggregator.
%	
%	\item 
%\end{enumerate}

%\subsection{Architecture}

%%\indent % indent 1st para also

\textit{We have designed a Cyber-Physical Anomaly Detection Engine for the aggregator that detects unexpected packets during communication}. This anomaly detection engine uses information from multiple sources to monitor the system as shown in Fig.\ref{fig::sysmodel}. Anomalous incoming packets to the aggregator are dropped thereby ensuring {\em intrusion tolerance} at the aggregator.
%The IDS consists of an Anomaly Detection Module 
%and a S3A Decision Module \cite{s3a}. 
%The Anomaly Detection Module monitors the system for anomalies and signals the S3A Decision Module when an anomaly is detected in the aggregator.
%The S3A Decision Module then switches control to a default trusted controller (aggregator in this case) and ensures safe system operation.
%\subsubsection{Decision Module}
%\indent % indent 1st para also

\begin{figure}[h]
	\centerline{\includegraphics[scale=0.28]{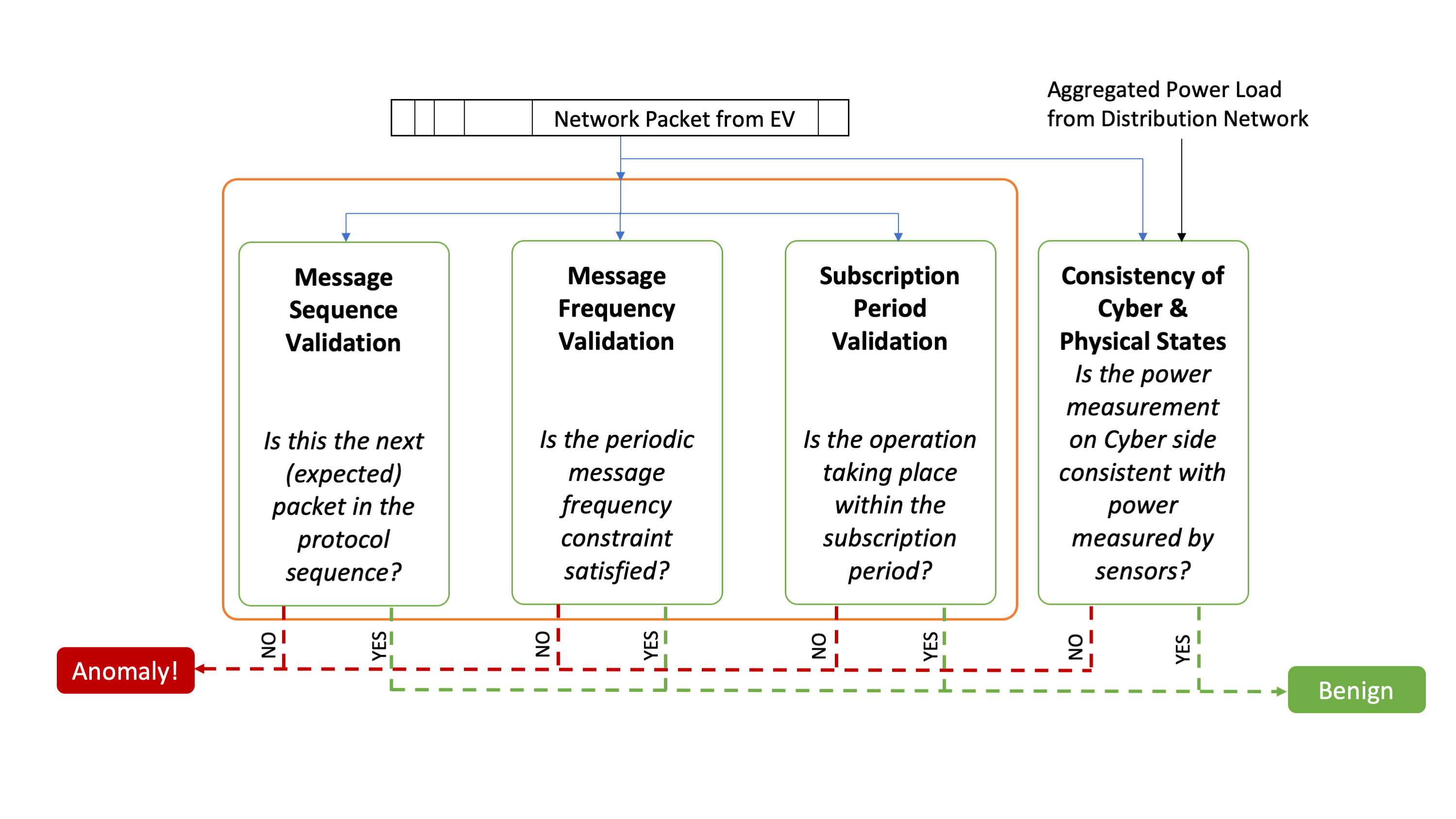}}
	\caption{Components of Cyber-Physical Anomaly Detection Engine}
	\label{fig::ids}
\end{figure}

The anomaly detection engine includes \ca message sequence validation \cb message frequency validation \cc subscription period validation and \cd  power measurement validation to detect anomalies in system behavior as shown in Fig.\ref{fig::ids}. An anomaly is detected whenever the system deviates from the expected system behavior. Expected system behavior is defined based on the communication standards in the V2G system, as discussed in detail below.

%\FloatBarrier
%\FloatBarrier
%\begin{itemize}
% \subsection{Cyber side Validations}
% %\begin{enumerate}
% In order to detect anomalies at the aggregator, data present in incoming packets from EVs as well as timing constraints associated with them are monitored. Two levels of deep packet inspection are used for this purpose as explained in detail below.

\subsection{Message Sequence Validation}
\label{subsec::ms}
%\indent
The sequences of messages in incoming packets are validated. To do this,
an aggregator state machine is created with valid states and state
transitions. This state machine is based on valid message sequences established by communication standards between the aggregator and the EVs as well as between the aggregator and the power grid. The
requirements and specifications for communication between
EVs and the electric power grid are established by the SAE
J2847/1 standard for forward power flow that includes charging \cite{sae, sae} and SAE J2847/3 standard
for reverse power flow that includes discharging 
% and frequency regulation 
\cite{sae}. 
A list of valid messages or commands between the aggregator and the EVs as well as the aggregator and the power grid is created.
Then a list of valid command sequences formed by these messages is created to verify if the packets coming into and going out of the aggregator follow these valid sequences. These valid messages and message sequences are in accordance with the SAE communication standards.

Since the SAE standards are proprietary, complete details are not provided. However, sufficient details on the types and sequences of messages are provided below in Fig. \ref{fig::agg} for a better understanding of our work. 

%\FloatBarrier
\begin{figure}[h]
	\centerline{\includegraphics[scale=0.25]{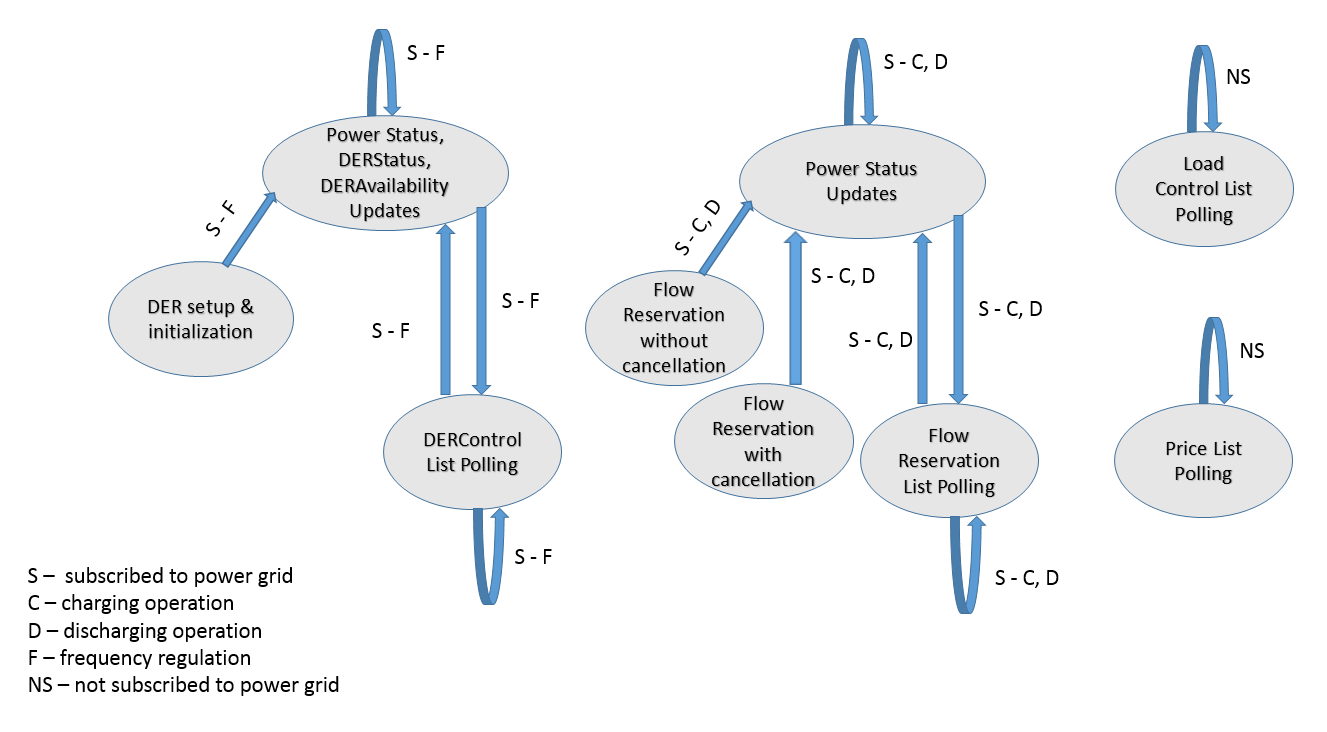}}
	\caption{Sub-states of the aggregator state machine.}
	\label{fig::agg}
\end{figure}
%\FloatBarrier

Fig. \ref{fig::agg} shows the various messages in each of the two EV-grid operations, \ca charging, \cb discharging.
% and \cc frequency regulation. 
It also shows the principal sub-states in the aggregator state machine that are generated from the list of valid command sequences enumerated for the three EV-grid operations. The state diagram captures the various states and state transitions when EVs are subscribed \ie connected to the power grid (denoted by 'S') as well as when they are not subscribed to the grid (denoted by 'NS'). When subscribed, EVs can be involved in any one of the three aforementioned EV-grid operations. Each state represents a \textit{sequence} of messages. The state transitions in various cases are as follows:

 \begin{enumerate}
 	\item [a)] Subscribed and Charging / Discharging - Flow reservation is the process in which the EV is assigned a subscription period (\ie period when connected to the grid) for charging/discharging. First, a flow reservation is established. This may be followed by one or more new flow reservations after cancellation of a previously established flow reservation. Once the subscription period begins, the EV starts sending periodic power status updates (\ie information related to the amount of power drawn) to the grid through the aggregator. In parallel, it also periodically fetches the flow reservation list from the grid, through the aggregator, to check for any updates in the subscription period.
%  	\item [b)] Subscribed and Frequency Regulation - first the set up and initialization of the Distributed Energy Resource (DER) is performed where the EV specific information (such as battery capacity, \etc) is sent and the process of frequency regulation is initiated. Once the subscription period begins, the EV starts sending periodic updates including power status, DER status and DER availability to the grid. In parallel, it also periodically fetches the DERControl list from the grid to check for any updates in the subscription period. The DERControl List is equivalent to the flow reservation list in case of charging/discharging operations and contains subscription period related information.
 	\item [b)] Not subscribed - when the EV is not subscribed to the power grid (\ie not engaged in any of the three EV-grid operations), it periodically fetches updates on pricing and load control related information from the grid to make a decision on when to charge / discharge.
 \end{enumerate}
 This component of the anomaly detection engine handles attack 3 in Tab.\ref{tab::threat}.
 
\subsection{Message Frequency Validation}
\label{subsec::mf}
%\indent
Power grid systems such as V2G system must satisfy certain time related constraints. Our anomaly detection engine monitors incoming packets to ensure that timing constraints are enforced. Periodic messages have a predefined frequency. The frequency of such periodic messages coming into the aggregator are monitored. At the aggregator, the frequency of a given periodic message is monitored by checking the time elapsed between two occurrences of the message. Therefore, it is not required that the clocks at the aggregator and the EVs be synchronized.
This component of the anomaly detection engine handles attack 5 in Tab.\ref{tab::threat}.
\subsection{Subscription Period Validation}
\label{subsec::sp}
%\indent

 \textit{Message data} in the packets coming into the aggregator are validated. Analysis of message data in packets is particularly useful for monitoring the aggregator/EV communication that involves a highly vulnerable edge device \viz the EV. Two important parameters in the EV to power grid communication that are most likely to be tampered by adversaries are:
\begin{enumerate}
	\item [a)] the subscription period, that defines the duration of charging/discharging during the respective
% 	/frequency regulation 
	operations and
	\item [b)] the state of charge (SOC), that defines the percentage of charge in the battery of the connected EV.
\end{enumerate}
Hence, we need to inspect packets to monitor these quantities. During charging and discharging operations, the EV periodically fetches the flow reservation list from the grid while also periodically updating its power status to the grid. 
% Similarly, during frequency regulation operation, the EV fetches the DERControl list from the grid. 
The flow reservation list 
% and the DERControl list 
contains the start and end of the subscription period. This data is used to verify that there are no power status updates outside the specified time interval. Power status updates occur only during the subscription period. This component of the anomaly detection engine handles attack 4 in Tab.\ref{tab::threat}.

The power status update messages contain vehicle SOC related information in terms of amount of power drawn. These power measurements are validated against physical power measurements as discussed next.

%\end{enumerate}
\subsection{Consistency of Cyber and Physical States}
\label{subsec::cp}
%\indent

%At this stage, we assume that the Physical State of the system is not compromised, \ie, the power measurements obtained from the network for validating the Cyber state of the system are genuine.
%We propose another level of security check by verifying the consistency of Cyber states with the Physical states. Power measurements are obtained from the power grid network through sensors \cite{pq} and compared with the power measurements reported in the cyber messages, \ie the power status update messages. In order to reduce the possibility of simultaneous tampering of both the cyber messages and the power measurements, the sensors used for cross-validation are located geographically far away from the EVs.
%\todo[inline]{@shanny: add details wrt end time being more noticeable in EVs with higher rated power (updated)}
We develop additional security checks to verify the consistency of cyber states with the physical states. Power measurements related to the EV charging load can be obtained from the sensors installed on the power distribution grid. Potential sources for the actual power data include  the connected EV charging equipment, smart meters and distribution line measuring devices. These sensors are assumed to be able to report the instantaneous active power or even complex power readings that are time-stamped for validating the cyber states of EV activities. 

\begin{figure*}[t!]
    \centering
    \begin{subfigure}[b]{0.5\textwidth}
        \centering
        \includegraphics[scale=0.45]{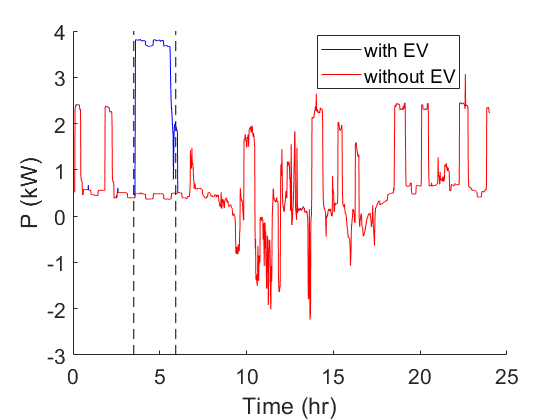}{a}
        \caption{Load profile of a residential home with and without the EV charging load.}
        \label{fig::aggregated_load}
    \end{subfigure}%
    ~ 
    \begin{subfigure}[b]{0.5\textwidth}
        \centering
        \includegraphics[scale=0.45]{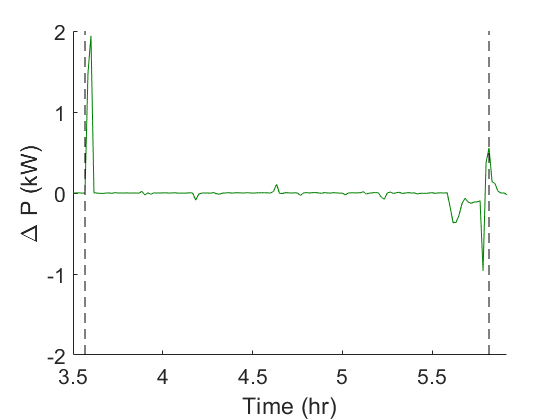}
        \caption{Filtered output of the total EV included load in Fig. \ref{fig::aggregated_load} during the EV charging period.}
        \label{fig::filtered_load}
    \end{subfigure}
    \caption{Power profiles}
\end{figure*}

To this end, let $P(t)$ denote the active power measurement for time period $t$, with the time resolution of $\Delta t$. We plot a sample daily profile of $P(t)$ for an actual residential home in Fig. \ref{fig::aggregated_load}, 
%\todo[inline]{@shanny: fix typo in fig - "withoug" (updated)}, 
both with and without the EV charging load. The residential home power demand and the EV charging data are obtained from the Pecan Street database at a minute-level sampling rate \cite{pecan_street}. 
As observed from the data, the EV charging load exhibits unique pattern that is different from other residential appliances and devices. First, the EV power demand is at least 3kW and also higher than that of other typical household loads \cite{7286621}. Based on the observed power profiles from the Pecan Street dataset, only the air-conditioning load has a comparable level of power demand. Second, the EV charging typically lasts for hours at the constant rated power demand level, which is different from the periodic pattern of air-conditioning load. Therefore, for the EV load, there is a noticeable change only at the start and end time points of charging. 

To better demonstrate this feature of EV load, we process the power data using a simple high pass filter to determine the rate of change between two consecutive data points, namely $\Delta P(t) = P(t)- P(t-\Delta t)$. To capture the fast change due to EV charging, the sampling rate of the power data needs to be sufficiently high to show that the EV can reach its rated power within two minutes while all other residential loads stay relatively unchanged. If the sampling rate increases, one may need to perform more sophisticated filtering process to recover the EV charging states. 

The filtered output for the residential home load with EV charging in Fig. \ref{fig::aggregated_load} from approximately 3:35 – 6:00 AM is illustrated in Fig. \ref{fig::filtered_load}. Note that the negative power in the aggregated load in Fig. \ref{fig::aggregated_load} is due to PV generation. The first spike in the plot represents the time instant when the EV charging started, reaching its rated power of around 3kW within two minutes. Note that when the charging ended for this 3kW rated EV, its power consumption slowly drops from the rated power and is therefore not noticeable from the filtered output. However, for EVs rated around 6kW or higher from the Pecan Street dataset, the end charging time is more noticeable from the filtered output. Specifically, the 'end charging' characteristics are very similar to the 'start charging' ones as the EV's power consumption drops from its rated power to zero within two minutes. This will result in a negative spike in the filtered output with a magnitude close to its rated power.

%\todo[inline]{@chaitra: For integration - to be confirmed} 
The power data spike due to start/end of EV charging can be used to verify the physical state when a packet with power measurement message is detected on the cyber network. Note that if the load profile is at sufficiently high sampling rate, it is unlikely that there is other major load change activity at the start/end time of EV charging. Therefore, the turn-on/-off events of air-conditioning loads will not confuse the engine with a potential EV activity. If the sampling rate gets slower, it will be necessary to smooth out the non-EV loads in order to determine the power spike from EV charging. Note that this may reduce the confidence in physical state verification part due to the existence of other heavy loading appliances with periodic patterns (\ie air-conditioning load). This component of the anomaly detection engine handles attacks 1,2 in Tab.\ref{tab::threat}.

We have performed a design space exploration for many combinations of EV and household numbers for the SAE protocol family (Fig. \ref{fig::sysmodel}). For instance,
\ci it can handle multiple EVs at a single household even if they start charging at the same instance of time. In which case, the total power consumed by these EVs is used to validate consistency on cyber and physical sides
\cii it can  can detect even if an EV stops charging partway before reaching 100\%

%\FloatBarrier

%% file: impl.tex
\section{Implementation}
\label{sec::impl}

To evaluate the anomaly detection engine, we implemented a prototype in Python2.7 on the Intel i7 NUC platform with
specifications as follows:
\begin{itemize}
    \item Platform - Intel i7 NUC
    \item Processor - Intel(R) Core(TM) i7-7567U CPU @ 3.5 GHz, 4 cores 
    \item Memory -  32 GB RAM, 128 GB HDD
    \item Operating System - Ubuntu 16.04
\end{itemize}
%\FloatBarrier
% \begin{table*}[h]
% 	\caption{System specifications}
% 	\begin{center}
% 		\renewcommand{\arraystretch}{1}
% 		\begin{tabular}{|p{2cm}|p{5cm}|p{3cm}|p{2.5cm}|}
% 			\hline
% 			\textbf{Platform} & \textbf{Processor} & \textbf{Memory} & \textbf{Operating System} \\
% 			\hline
% 			 Intel i7 NUC & Intel(R) Core(TM) i7-7567U CPU @ 3.5 GHz, 4 cores & 32 GB RAM, 128 GB HDD & Ubuntu 16.04 \\
% 			\hline
% 		\end{tabular}
% 		\label{tab::spec}
% 	\end{center}
% \end{table*}
%\FloatBarrier

The algorithm for anomaly detection is as shown in Algorithm \ref{ade} in Appendix. The engine inspects all the received packets from EVs and filters out the malicious ones.
To handle multiple EVs at an aggregator, there are multiple instances of packet inspection that execute in parallel, each instance monitoring a single EV.
% The maximum number of such parallel instances is same as the maximum number of EVs that can be handled by a given aggregator.
The source EV of each received packet is first identified based on its IP address and the packet is then passed onto the corresponding instance of packet inspection (Algorithm \ref{pid} in Appendix). Based on the output from the packet inspection module, the packet is either dropped or output. The algorithm for packet inspection as detailed in \S \ref{sec::ids} is shown in Algorithm \ref{pins} in Appendix.
% The algorithm for packet identification is shown in Algorithm \ref{pid} in Appendix. The source and destination IP addresses of a packet are used to determine the instance i to which the packet must be passed on. If the packet is from a new EV, the packet is either assigned to a free instance (an instance i in idle state) or dropped if all instances are busy (\ie -1 is returned). An instance moves to idle state if it does not receive a packet from the EV for a specified time period (i.e., connection to the EV is dropped after a timeout).
%The python scapy module \cite{scapy} is used for packet parsing. 
% Using the specifications provided by the SAE communication standards for forward and reverse power flows \cite{sae},  a state machine of all possible valid states and state transitions is created. The received packet is parsed and \ca message sequence validation, \cb message frequency validation, \cc subscription period validation and \cd cyber-physical state consistency validation using power measurements from physical sensors embedded in the network are performed to detect anomalous behavior (as discussed in detail in \S \ref{sec::ids}). Based on the outcome of this inspection, the state machine is advanced and anomaly notifications are returned accordingly. As previously mentioned, an instance of the packet inspection module moves to idle state if no packet is received until timeout.

Algorithm \ref{pv} in Appendix is used to validate the consistency of the cyber and physical states by detecting anomalies in the EV charging. To detect the EV chrg state from just the aggregated load perspective, we first pass the aggregated load through a high pass filter described in Algorithm \ref{hpf} in Appendix. Next, Algorithm \ref{ev_csi} in Appendix is used to determine the EV 'chrg state' by searching for a $\Delta P$ within an acceptable range of EV rated power that corresponds to a 'start charging' or 'stop charging' event.

%% file: eval.tex
%\newpage
%\setcounter{figure}{0}
%\setcounter{table}{0}
\section{Evaluation}
\label{sec::eval}

Our focus is to secure the aggregator, a critical component of the V2G system, thereby increasing
the reliability and the resiliency of such systems against cyber attacks. Power grid systems are often time-sensitive. Therefore, our goal is to not only detect anomalies
accurately but to also do so as soon as a malicious network packet arrives at the aggregator. The anomaly detection engine is placed at the aggregator in the V2G system as shown in Fig. \ref{fig::sysmodel}. This makes it important to ensure that it does not introduce significant delay to the packet transfer rate at the aggregator. We evaluate the prototype of our anomaly detection engine for both accuracy and performance: 
\begin{enumerate}
	\item \textbf{Accuracy} is measured in terms of \textit{false positives} and \textit{false negatives}.
	
	\item \textbf{Performance} is measured in terms of \textit{average} and \textit{ worst case} time taken by the anomaly detection engine to inspect one packet and compared with minimum network latency.
\end{enumerate}
Note that as mentioned previously, since there are no existing anomaly detection engines to the best of our knowledge that combine cyber and physical properties along with communication standards for this type of system (\S{\ref{sec::rel_work}}), direct comparisons are not possible.

\textbf{Power Data for EV Charging} - Packets on the cyber network are generated based on EV power profile data for charging obtained from the Pecan Street database at a minute-level sampling rate \cite{pecan_street}. This data consists of timestamps and corresponding power measurements with respect to each EV, from which all required information (such as subscription period) for packet generation can be extracted.

\textbf{Power Data for EV Discharging} - Using the above EV charging data, we formulated the data for EV discharging based on efficiency formulas from paper \cite{dischrg_data}. From equation (12) in the reference paper \cite{dischrg_data}, the magnitude of EV charging/discharging power from/to the
power grid is described as follows

\begin{equation}
  |P_{grid}|=\begin{cases}
    \dfrac{EV_{rated}}{\eta_1} = |P_c|\\
    EV_{rated}*{\eta_2} = |P_d|
  \end{cases}
\end{equation}

$EV_{rated}$ is the rated EV power, $\eta_1$ is charging efficiency, $\eta_2$ is discharging efficiency, $|P_c|$ is the magnitude of EV charging power and $|P_d|$ is the magnitude of EV discharging power. Using algebraic manipulation and substituting $EV_{rated}$, we obtain the magnitude of EV discharging power as a function of EV charging power.
From the reference paper \cite{dischrg_data}, we
obtain the charging and discharging efficiencies in Table II as 0.92 and 0.92 respectively. Plugging these values into the previous equation, we get the magnitude of the EV discharging power is approximately 85\% of the EV charging power.

\begin{equation}
   |P_d| = |P_c| * {\eta_1} * {\eta_2} = |P_c| * 0.92 * 0.92 = |P_c| * 0.846
\end{equation}

% We append an extra binary attribute, 'chrg state' = 1 (charging) or 0 (not charging) to aid in the extraction of subscription periods of operations for packet generation. The EV chrg state = 1 when the EV is in the start charging, steady state charging, and stop charging phases. Otherwise, the EV chrg state = 0 as it is not charging. 
% The corresponding aggregated power profile data (that includes EV load along with other residential loads) is used for cross-validation on the physical side.

Currently, there are no EVs/EVSEs that support the SAE J2847/1 and SAE J2847/3 standards since these communication standards are still in the process of development. There has been significant effort towards making the real world implementation of SAE standards feasible \cite{saeimp1,saeimp2}. Once these standards are implemented, our anomaly detection engine can be easily integrated with them.
Therefore, to test our anomaly detection engine, we %developed a python script to 
generate packets with custom HTTP payloads according to specifications provided by SAE communication standards\cite{sae}.
%\footnote{We wrote a Python script for this purpose.}.
The simulation is carried out on an Intel i7 NUC (\S \ref{sec::impl}). 
% Packets are generated and written to a pcap file by the packet generator script. These packets are then read from the pcap file and validated by the anomaly detection engine script. 
Further details on how the packets are generated are provided below.

\subsection{Accuracy}
\label{subsec::acc}
Testcases to test for false positives are generated based on certain ground truth and testcases to test for false negatives are obtained by modifying the former testcases to introduce various kinds of anomalies.

\textbf{Message Sequence Validation} - The ground truth for testcase generation here is the SAE standards \cite{sae}. \ca Testcases for false positives include all possible valid sequences consisting of parallel as well as repeating sequences. Due to the possibility of a lot of valid variations, there is a large number of these test cases. For instance, consider the charging operation. As shown in Fig. \ref{fig::agg}, first a flow reservation with or without cancellation is performed (note the existence of two possibilities already). Then the EV starts sending periodic power updates to the grid during its subscription period for charging. In parallel, the EV also periodically fetches the flow reservation list from the grid. Periodic messages give rise to repeating sequences and increase the number of possible valid variations. Similarly, parallel sequences of messages (power updates and fetching of flow reservation list in this case) also increase the number of possible valid variations. This is because one or more messages from a parallel sequence (say, power updates in this case) can arrive anywhere between messages in a related parallel sequence (fetching of flow reservation list in this case). The sequence to which the message belongs is identified using the message data. \cb Testcases for false negatives were generated by randomly placing invalid packets amidst valid sequences.

\textbf{Message Frequency Validation} - The ground truth for testcase generation here is again the SAE standards \cite{sae}. \ca Testcases for false positives consisted of packets with expected message periodicities. \cb Testcases for false negatives consisted of packets containing messages with periodicities different (\ie periodicities lower and higher than expected values) from expected values as specified in the SAE standards \cite{sae}. 

\textbf{Subscription Period Validation} - The ground truth here is based on the data obtained from the Pecan Street database \cite{pecan_street}. \ca Testcases for false positives consisted of packets with subscription periods consistent with the aforementioned data. \cb Testcases for false negatives consisted of packets with inconsistencies where the actual subscription period was different from previously agreed upon subscription period, i.e, the arrival time of packets containing power status updates were modified so as to be different from the time intervals specified in packets containing the flow reservation list.

\textbf{Consistency of Cyber and Physical States} - We have a high certainty on detecting the stop charging time for EVs with higher rated power versus EVs with lower rated power for two reasons. The first is that the stop charging characteristic for higher rated EVs is very similar to the start charging characteristics of EVs except that it will drop from its rated power to zero within two minutes. The lower rated EVs take longer than two minutes to stop charging so there will not be a noticeable negative spike in the filtered load sequence. The second reason is that the higher rated EVs are less sensitive to non-EV loads changes. For instance, if an 1.5kW AC unit starts/stops around an EV start/stop event, it will effect the total $\Delta P$ of two consecutive time stamps. In the worst case scenario, the AC will reach its rated power within one minute. The AC rated power is 50 percent of an EV rated at 3kW which will force $\Delta P$ out of the acceptable range of start/stop charging values. The AC rated power is only 25 percent of an EV rated at 6kW, however, $\Delta P$ will usually be barely within range of acceptable start/stop charging values making higher rated EVs less sensitive to non-EV load changes. The ground truth here is again based on the power data obtained from the Pecan Street database \cite{pecan_street}. \ca Testcases for false positives consisted of packets with power measurements consistent with this data. During steady state charging, the EV power should not vary beyond +/- 0.5kW in the worst case scenario. Any power measurement beyond this range is considered anomalous. \cb Testcases for false negatives makes use of this fact, \ie it consists of packets with inconsistencies with respect to power measurements, i.e, power measurements were varied to be above and below this permissible range of +/- 0.5kW.

\subsection{Performance}
%\indent % indent 1st para also

\textbf{Minimum network latency} - According to the smart grid communication requirements specified by the Department of
Energy, the minimum network latency with reference to Electric Transportation is \textit{2 seconds} \cite{doe}

% According to the smart grid communication requirements specified by the Department of
% Energy, the maximum network bandwidth with reference to Electric
% Transportation is 56 kbps \cite{doe}. The packet size in this application is approximately 0.5 KB, being computed over the set of messages defined in SAE standards. Thus, the maximum network bandwidth in terms of number of packets
% is approximately 14 packets per second and the minimum inter-arrival time between two packets on the network is 0.0714 seconds.

\textbf{Worst case time taken} - We compare the worst case time taken by the anomaly detection engine to inspect a packet with the minimum network latency to determine whether or not the anomaly detection engine introduces significant delay. The worst case time taken to inspect a packet is when it goes through the longest datapath in the anomaly detection engine. The 99.9th percentile worst case time is \textit{0.165 seconds}.

\textbf{Average time taken} - In this evaluation, the total number of EVs that are simultaneously handled by the aggregator has been varied (up to a maximum of 400) based on literature \cite{ev_load}.
\begin{figure}
  \begin{center}
    \includegraphics[scale=0.3]{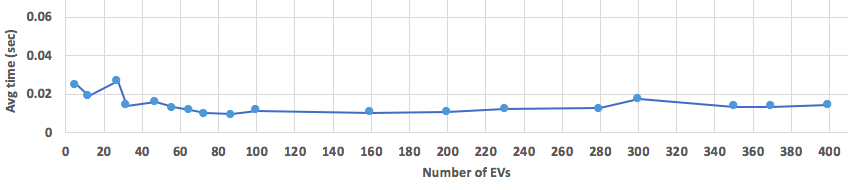}
  \end{center}
  \caption{Performance of the Anomaly Detection Engine}
  \label{fig::plot}
\end{figure}
As shown in Fig. \ref{fig::sysmodel}, there are multiple households connected to a single aggregator with (possibly) multiple EVs connected to each household. The power measurements originally obtained from the Pecan Street database \cite{pecan_street}) contains individual EV loads as well as the corresponding aggregated loads at households where one household has just one EV associated with it. We have modified this power data as follows:
\ci In order to simulate multiple EVs connected to each household,
we first assign
% we used a function to modify the aggregated power load of the household as well as the individual EV power loads. The function takes as input
%the desired house to generate data for, the desired EVs to be connected and their
charging/discharging start times to include situations where multiple EVs at one household start to charge/discharge either at the same instance of time or at different instances of time. Then, we compute the operation end times while keeping the duration of operation unchanged. Note that we are only modifying the charging/discharging start times for EVs  \ie we are just sliding the EV charging/discharging duration windows (Algorithm \ref{algo::load_gen} in Appendix).
 %the desired house's grid data with no EV, and the charging data for all EVs. 
% The function then aggregates the household power load (without any EVs) with the modified power load for each individual EV to generate the modified aggregated power load of the household. %charging data at the desired start charging times. 
%It then outputs the aggregated load data and each individual EV's charging profile into a csv file.
\cii To simulate multiple such households being simultaneously connected to the aggregator, we replicated the power data per household. The average time taken by the anomaly detection engine to inspect a packet is approximately \textit{0.014 seconds} as shown in Fig. \ref{fig::plot}.
Tab. \ref{perf} summarizes these results.
\linespread{0.9}
%\FloatBarrier
\begin{table}[h]
	\centering
	\caption{Performance of Anomaly Detection Engine}\label{perf}
	\renewcommand{\arraystretch}{1}
	\begin{tabular}{|p{1.5cm}|p{2.5cm}|p{2.5cm}|}
		\hline
		\textbf{Minimum Network Latency} & \textbf{Worst case per packet inspection time} & \textbf{Average per packet inspection time} \\
		\hline
		2 seconds & 0.165 seconds & 0.014 seconds \\
		\hline
	\end{tabular}
\end{table}
%\FloatBarrier
\linespread{1}

As stated previously (\S \ref{sec::system}), we assume that the cyber network connecting physical network sensors to the anomaly detection engine has similar properties (in terms of bandwidth etc) as the cyber network in the V2G system. Therefore, the individual EV power measurements on the V2G cyber network and the corresponding aggregated power measurements on the sensor cyber network are received by the anomaly detection engine simultaneously (\ie without any delay).

% %\FloatBarrier
% \begin{figure*}[h]
% 	\centerline{\includegraphics[scale=0.5]{images/plot.png}}
% 	\caption{Performance of the Anomaly Detection Engine}
% 	\label{fig::plot}
% \end{figure*}
% %\FloatBarrier

%% file: disc.tex
\section{Discussion}
\label{sec::disc}

As seen from the results in Tab. \ref{perf}, the worst case time to inspect a packet \ie \textit{0.165 seconds} is lesser than the minimum network latency, \ie 2 seconds. 
% So is the
% average time for inspection of a packet, \ie \textit{0.014 seconds}.
Therefore, our anomaly detection
engine detects malicious packets almost instantly without
introducing significant delay at the aggregator.
% We have tested for false positives and false negatives to validate the accuracy of our prototype. The lack of false positives and false negatives merely show that the state machine is correctly implemented.

In all of our validation techniques, the consistency of data obtained from one source is verified against data obtained from a different source as explained next. This makes it more difficult to tamper with the data so as to produce false consistencies.
In case of Message Sequence Validation and Message Frequency Validation,  packets are verified against information obtained from the communication standards documentation. For validating consistency of Cyber and Physical States, power data reported in cyber messages is validated against physical power measurements. In case of Subscription Period Validation, subscription period related data is validated using messages from the power grid side of communication that contain flow reservation list
% /DERControl List 
and messages from the EV side of communication that contain power status updates.

Packet loss can result in certain packets either arriving (at the aggregator) out of sequence or at unexpected times. This might result in false positives being signalled by the anomaly detection engine. However, rate of packet loss is very low given the 99-99.99\% network reliability requirement for Electric Transportation in the Smart Grid \cite{doe}.
Network delays/jitters could also lead to false positives due to packets arriving at an unexpected time. This can be handled by setting an appropriate tolerance level with respect to time constraints.
Our anomaly detection engine has a complete list of all possible valid states and state transitions. This eliminates the likelihood of false negatives.
Our evaluation is carried out based on the previously stated assumption that the inputs obtained from atleast one of the cyber or physical networks are genuine. If this is not true, then an attacker could tamper with the cyber and physical states of the system to obtain consistency and conceal the attack from our anomaly detection engine.

%If the aggregator/anomaly detection engine restarts unexpectedly (and it is assumed that all of the previous state information is lost) while some EVs are actively connected to the grid, then the effect on various components of the anomaly detection engine is as follows. \ca \textbf{Consistency of Cyber and Physical States} - this verification continues unhindered, \ie, inconsistencies in power measurements can be detected instantaneously from the time the engine restarts. \cb \textbf{Message Sequence Validation} - instantaneous detection of sequence related anomalies after restart is not possible until the state machine receives enough packets to learn the current EV-grid operation. This is due to the fact that many EV-grid operations have overlapping sub-sequences in their communication messages. \cc \textbf{Message Frequency Validation} - excluding the first periodic message packet arriving after restart, frequency validation can be performed on all the following periodic message packets. \cd \textbf{Subscription Period Validation} - if we assume that the packets containing subscription period related details arrived at the anomaly detection engine before restart and the engine has lost this state information after restart, subscription period validation is not possible for the ongoing EV connection. However, this validation proceeds unhindered \ie instantaneously if the subscription period related packets arrive after engine restart.

%% file: rwork.tex
%\newpage
%\setcounter{figure}{0}
%\setcounter{table}{0}
\section{Related Work}
\label{sec::rel_work}

There has been a wide range of attacks on the power grid that has triggered research towards securing its various components such as the advanced metering infrastructure (AMI) and smart inverters. Examples of intrusion detection systems for cyber-physical systems include the work on securing the advanced metering system by using specification based intrusion detection \cite{t3}. This IDS monitors the cyber state of the system by observing traffic among access points and meters at various layers to ensure expected behavior. We use similar techniques to monitor the cyber state of a V2G aggregator. In addition, we also check for consistency of cyber and physical states of the system.

Apart from the various components of the power grid, the distribution networks have also received attention. The paper by Liao et al \cite{msp2} focuses on enhancing power grid security by using micro-synchrophasors as a tool to monitor and manage distribution networks. The high fidelity, time-synchronized phase angle and voltage
magnitude data obtained from micro-synchrophasors helps track events originating at local distribution. This work is similar to our work in that it uses data from sensors for monitoring. However, in addition to using sensor readings, our anomaly detection engine also monitors other data constraints related to the communication protocol specifications along with timing constraints related to message frequencies and subscription periods. 

With increase in number of DERs being integrated with the power grid, there has been effort in the direction of securing these DERs \cite{der}. The paper discusses 
the architecture of cyber-physical power system with penetration of DERs, analyzes related
cyber security challenges, summarizes important attack scenarios and proposes a DER resilience analysis methodology to prevent, detect and respond to attacks. The paper provides a generalized analysis for DERs. However, specific DERs have their own challenges. Our work focuses on a specific DER, i.e., EVs.

With respect to the V2G system, implementation and optimization
have so far received a lot of focus. The paper by Guille et al \cite{vi} discusses a framework to integrate EVs with the power grid. The paper by Mal et al \cite{ec} focuses on optimizing the charging operation in the V2G system and thereby efficiently balancing the load on power grid. However, attention has recently shifted towards the security of EVs in the power grid. Chen et al \cite{an} propose an efficient and secure
authentication scheme for V2G networks that preserves privacy. The authentication scheme provides anonymity in the
V2G network.
The paper focuses on securing the communication of EVs in the V2G system of power grid. 
On the other hand, our work focuses on securing the aggregator, an important component of the V2G system, by increasing its resiliency to attacks.

There is some related work on identifying EV charging profiles for improving power distribution system operations. The statistical characteristics of EV's state-of-charge or the duration of charging period have been studied in \cite{su2012performance,qian2011modeling} by analyzing a fleet of EV charging profiles. More recently,  a deep learning approach has been proposed in \cite{wang2018robust} to extract the EV charging profile from the aggregated household demand as a load disaggregation problem. However, we use a similar filtering mechanism but for a different purpose, \ie anomaly detection to ensure security.
We have developed a simple approach for estimating EV charging status that is very suitable for real-time implementation needs of the proposed anomaly detection engine. Interesting future work opens up on investigating the potential trade-offs between accuracy and computational complexity for the task of determining EV charging activity from aggregated load data.

%% file: concl.tex
%\newpage
%\setcounter{figure}{0}
%\setcounter{table}{0}
\section{Conclusion}
\label{sec::concl}

%  \hl{A conclusion is NOT a summary of the paper. A conclusion should include some previously undisclosed nuggest/insight -- either how the system can be used or adapted or so. A high level description of the paper (1-2 sentences is fine) is OK but not this level of summary. Many of these sentences are repeats from the paper.}
In this work, we have presented a novel architecture of a Cyber-Physical Anomaly Detection Engine that captures the cyber and physical properties of the system along with the related communication standards to define correct system behavior. 
% It is designed for the aggregator which an important component of the Vehicle-to-Grid System that integrates Electric Vehicles with the Power Grid.
% As discussed previously, in all the validation techniques used for anomaly detection, consistency of data obtained from one source is verified against data obtained from a different source, thus making it difficult for adversaries to tamper with these data so as to produce false consistencies.
The simple model of our anomaly detection engine demonstrates that accurate and almost instantaneous detection of anomalies is feasible.
Although our prototype Cyber-Physical Anomaly Detection Engine is based on SAE standards of communication for V2G system, this architecture can be extended to other communication standards for other DERs as well.

%% file: ack.tex
%\newpage
\setcounter{secnumdepth}{0}
\section{Acknowledgements}

This work is supported in part by grants from the Dept. of Energy (CEDS program - contract number DE-0E0000826) and the National Science Foundation (NSF SaTC 1718952). Any opinions, findings, and conclusions or recommendations expressed here are those of the authors and do not necessarily reflect the views of sponsors.
The authors would also like to thank Tim Yardley, Steve Granda, Devu Shila, Lynn Ren for their useful discussions and ideas.

%% file: app.tex
\newpage
\setcounter{secnumdepth}{0}
\section{Appendix}

\linespread{0.85}
%\FloatBarrier
% Insert the algorithm
\begin{algorithm}[H]
	\caption{EV Chrg State Identification}\label{ev_csi}
	\begin{algorithmic}[1]
		\STATE \textbf{Input:} Output sequence of \textbf{High\_Pass\_Filter}, EV rated power 
		\STATE \textbf{Output:} EV charge state sequence
		\STATE //\textit{Determine initial chrg state from EV load profile: 1 - charging, 0 - not charging}
		\IF{High EV rated power}
		\IF{EV load power is within steady state power range}
		\STATE set initial chrg state to 1
		\ELSE
		\STATE set initial chrg state to 0
		\ENDIF
		\FOR{i in filtered load sequence}
		\STATE //\textit{EV is currently not charging}
		\IF{previous chrg state = 0}
		\STATE //\textit{Look for EV charging event}
		\IF{current filtered data point $>$ 0}
		\STATE $\Delta P$ = current filtered load point + next filtered load point
		\IF{$\Delta P$ is within an acceptable range of EV rated power}
		\STATE //\textit{EV charging event detected}
		\STATE set chrg state to 1
		\ELSE
		\STATE set chrg state to 0
		\ENDIF
		\ENDIF
		\ELSE
		\STATE //\textit{previous chrg state = 1, look for stop charging event}
		\IF{current filter data point $<$ 0}
		\STATE $\Delta P$ = current filtered load point + previous filtered load point
		\IF{$\Delta P$ is within acceptable negative range of EV rated power}
		\STATE //\textit{EV stop charging event detected}
		\STATE set chrg state to 0
		\ELSE
		\STATE set chrg state to 1
		\ENDIF
		\ELSE
		\STATE set chrg state to 1
		\ENDIF
		\ENDIF
		\ENDFOR
		\ELSE
		\STATE //\textit{Lower EV rated power}
		\FOR{i in filtered load sequence}
		\IF{current filter data point $>$ 0}
		\STATE $\Delta P$ = current filtered load point + next filtered load point
		\IF{$\Delta P$ is within an acceptable range of EV rated power}
		\STATE //\textit{EV charging event detected}
		\STATE //\textit{Stop charging event cannot be detected - set a constant charging time period}
		\FOR{t in constant charging time period}
		\STATE set chrg state to 1
		\ENDFOR
		\ELSE
		\STATE set chrg state to 0
		\ENDIF
		\ENDIF
		\ENDFOR
		\ENDIF
	\end{algorithmic}
\end{algorithm}
%\FloatBarrier
\linespread{1}

%\linespread{0.4}
%\FloatBarrier
\begin{algorithm}[h]
	\caption{Cyber-Physical Anomaly Detection Engine}\label{ade}
	\begin{algorithmic}[1]
		\STATE \textbf{Input:} All packets from EVs
		\STATE \textbf{Output:} Benign packets from EVs
		\STATE Spawn 'n' instances of \textbf{Packet\_Inspection} module
		\FOR{packet in packets}
		\STATE instance\_i := \textbf{Packet\_Identification}\big(packet\big)
		\IF{ instance\_i $\neq$ -1}
		\STATE //\textit{Call ith instance of Packet\_Inspection}
		\STATE anomalous := \textbf{Packet\_Inspection}\big(packet\big)
		\IF{anomalous}
		\STATE drop packet
		\ELSE
		\STATE output packet
		\ENDIF
		\ELSE
		\STATE drop packet
		\ENDIF
		\ENDFOR
	\end{algorithmic}
\end{algorithm}
%\FloatBarrier
%\FloatBarrier
% Insert the algorithm
\begin{algorithm}[h]
	\caption{Packet Identification}\label{pid}
	\begin{algorithmic}[1]
		\STATE \textbf{Input:} Packet
		\STATE \textbf{Output:} Instance of \textbf{Packet\_Inspection} module
		\STATE //\textit{Determine to which instance of Packet\_Inspection module does packet belong}
		\FOR{i in instances}
		\STATE //\textit{Checking src and dst IP addresses}
		\IF{packet IP equals IP monitored by i} 
		\STATE return i
		\ELSE
		\IF{i is free}
		\STATE set monitoring IP of i to packet src IP and return i
		\ELSE
		\STATE return -1
		\ENDIF
		\ENDIF
		\ENDFOR
	\end{algorithmic}
\end{algorithm}
%\FloatBarrier
%\FloatBarrier
% Insert the algorithm
\begin{algorithm}[h]
	\caption{Packet Inspection}\label{pins}
	\begin{algorithmic}[1]
		\STATE \textbf{Input:} Packet
		\STATE \textbf{Output:} Packet is anomalous or not
		\IF{no packet received until timeout}
		\STATE transition to idle state
		\ELSE
		\STATE //\textit{Message Sequence Validation}
		\IF{packet.payload matches expected packet.payload in current state}
		\STATE //\textit{Message Frequency Validation}
		\IF{(packet.arrivalTime - packet.lastArrivalTime) satisfies frequency constraints}
		\STATE update packet.lastArrivalTime
		\ELSE
		\STATE raise anomaly - inconsistent frequency
		\ENDIF
		\STATE //\textit{Subscription Period Validation}
		\IF{current time in subscription period and packet.payload contains power status update}
		\STATE pass
		\ELSE
		\STATE raise anomaly - invalid subscription
		\ENDIF
		\STATE //\textit{Cyber and Physical States Consistency Validation}
		\STATE match:=\textbf{Power\_Validator}\big(packet.power, packet.timestamp, packet.srcIP\big)
		\STATE //\textit{if power reported in packet matches filtered EV power from aggregated load power}
		\IF{match is True}
		\STATE pass
		\ELSE
		\STATE raise anomaly - inconsistent power
		\ENDIF
		
		\STATE transition to next state
		
		\ELSE
		\STATE raise anomaly - unexpected packet
		\ENDIF
		\ENDIF
		\IF{anomaly raised} 
		\STATE return anomalous
		\ELSE
		\STATE return not anomalous
		\ENDIF
	\end{algorithmic}
\end{algorithm}
%\FloatBarrier
%\FloatBarrier
% Insert the algorithm
\begin{algorithm}[h]
    \caption{High Pass Filter} \label{hpf}
    \begin{algorithmic}[1]
        \STATE \textbf{Input:} Aggregated load sequence
        \STATE \textbf{Output:} Filtered aggregated load sequence
        \FOR{i in aggregated load sequence}
        \STATE i in filtered aggregated load sequence = i in aggregated load sequence + i-1 in aggregated load sequence
        \ENDFOR
    \end{algorithmic}
\end{algorithm}
%\FloatBarrier
% Insert the algorithm
\begin{algorithm}[h]
    \caption{Power Validator}\label{pv}
    \begin{algorithmic}[1]
        \STATE \textbf{Input:} packet.Power, packet.timestamp, packet.srcIP
        \STATE \textbf{Output:} True/False - Cyber and Physical States Consistency Validation
        \STATE //\textit{chrg state: 0 - not charging, 1 - charging}
        \STATE //\textit{Determine max and min steady state charging power for packet.srcIP}
        \STATE //\textit{Determine chrg state from output of \textbf{EV\_Chrg\_State\_Identification} at packet.timestamp for packet.srcIP}
        \IF{chrg state is 0}
        \STATE //\textit{No EV charging for given packet.timestamp}
        \STATE return 0
        \ELSE
        \STATE //\textit{chrg state is 1}
        \IF{packet.power $<$ min steady sate charging power for packet.srcIP}
        \IF{this is the first packet sent}
        \STATE //\textit{EV could be in the process of reaching its rated power}
        \STATE return 1
        \ELSE
        \STATE return 0
        \ENDIF
        \ELSIF{packet.power is within the range of max and min steady state power for packet.srcIP}
        \STATE return 1
        \ELSE
        \STATE return 0
        \ENDIF
        \ENDIF
    \end{algorithmic}
\end{algorithm}
%\FloatBarrier
%\FloatBarrier

%\FloatBarrier
% Insert the algorithm
\begin{algorithm}[h]
    \caption{Household Load Generator} 
    \label{algo::load_gen}
    \begin{algorithmic}[1]
        \STATE \textbf{Input:} household power profiles without EVs, individual EV power profiles, EV charging start times
        \STATE \textbf{Output:} new household aggregated power profiles including EVs, new individual EV power profiles 
      % \STATE Initialize output variables
        \FOR{all EVs}
        \IF{i-th EV is chosen to be included in household}
        \FOR{all charging events of EV}
        \STATE new EV charge end time = new EV charge start time + (old EV charge end time - old EV charge start time)
        \ENDFOR
        \STATE generate aggregated household power profile by adding new EV power profile to household power profile without EVs
        \ENDIF
        \ENDFOR
    \end{algorithmic}
\end{algorithm}